\documentstyle[11pt]{article}
\begin{document}
\begin{flushright}
CUPP-00/3 \\
hep-ph/0006260
\end{flushright}
\vskip 100pt
\begin{center}
{\Large {\bf Decay and Decoupling of heavy Right-handed Majorana
Neutrinos in the L-R model}}
\vskip 25pt
{\sf Paramita Adhya $^{a,\!\!}$
\footnote{E-mail address: paro27@hotmail.com}},
{\sf D. Rai Chaudhuri $^{a,\!\!}$
\footnote{E-mail address: allrc@usa.net}},
and
{\sf Amitava Raychaudhuri $^{b,\!\!}$
\footnote{E-mail address: amitava@cubmb.ernet.in}}
\vskip 10pt
$^a${\it Department of Physics, Presidency College, 86/1, College
Street,\\
Calcutta 700073, India}\\
$^b${\it Department of Physics, University of Calcutta, 92
Acharya Prafulla Chandra Road, \\ Calcutta 700009, India}\\
\vskip 20pt
{\bf Abstract}
\end{center}

{\small
Heavy right-handed neutrinos are of current interest. The
interactions and decay of such neutrinos determine their
decoupling epoch during the evolution of the universe. This in
turn affects various observable features like the energy density,
nucleosynthesis, CMBR spectrum, galaxy formation, and
baryogenesis. Here, we consider reduction of right-handed
electron-type Majorana neutrinos, in the left-right symmetric
model, by the $W_R^+W_R^-$ channel and the channel originating
from an anomaly, involving the $SU(2)_R$ gauge group, as well as
decay of such neutrinos. We study the reduction of these neutrinos
for different ranges of left-right model parameters, and find that,
if the neutrino mass exceeds the right-handed gauge boson mass,
then the neutrinos never decouple for realistic values of the
parameters, but, rather, decay in equilibrium. Because
there is no out-of-equilibrium decay, no mass bounds can be set
for the neutrinos.
}

\vskip 30pt

\begin{center}
\end{center}

\newpage
\section{Introduction}

If a see-saw mechanism \cite{gel} is to account for left-handed
electron neutrino $\nu$ masses sufficiently small to be
consistent with current ideas on neutrino oscillations,
right-handed $N$ neutrinos with mass, $M$, in the TeV scale come into
the picture \cite{14,maa}. High $N$ masses in the range 1-20 TeV,
and even higher masses, have been considered in studies of
leptogenesis -- baryogenesis \cite{14,fuku,pil} and $e^-e^-$
collisions \cite{maa}. While some studies \cite{pil} consider $M$
smaller than $M_W$, the right-handed $W_R$ boson mass, others
specifically use $N$ masses greater than the $W_R$ mass
\cite{maa,and}. As cosmological and laboratory lower bounds for
$Z'$, $W_R$ masses are of the order of 0.5 TeV, 1.6-3.2 TeV,
respectively \cite{17}, there is no reason not to consider $N$
masses greater than $Z'$, $W_R$ masses. Cosmological mass bounds
for $M$, with $M\gg M_W$, have been considered in
\cite{4}-\cite{pa}.

In a recent work \cite{pa}, $(B+L)$--violation from an anomaly,
involving the $SU(2)_R$ gauge group \cite{14,15,16}, was
considered as a generation/reduction channel for $N$ neutrinos
satisfying $M > M_W$. It was found that this anomalous channel
played a role, in the decoupling of such neutrinos, at least as
important as the $N\bar N\to W_R^+W_R^-$ channel (each of these
channels was found more important than the $N\bar N\to F\bar F$
channel, $F$ representing a relevant fermion). Matrix elements
for $N\bar N\to F\bar F$ and $N\bar N\to W_R^+W_R^-$ were
calculated in \cite{pa} from the Left-Right symmetric extension
\cite{9,10} of the standard model, as an illustration.

In the above work \cite{pa}, the $N$ neutrinos were assumed to be
stable, for simplicity. If, however, the Left-Right symmetric
model (L-R model, hereafter) is to be taken as a serious working
basis, $N$ neutrinos cannot be considered to be stable. Decays
involving $\nu-N$ mixing, $W_L-W_R$ mixing, and generation
mixing, and CP-violating decays have been extensively studied.
The last scenario has been widely used in generating lepton and,
hence, baryon number from decay of Majorana-type $N$ neutrinos
\cite{fuku,pil,luty,vay,flan}. If the $N$ mass is considered to
be greater than the $W_R$ mass, then these decay channels will be
marginalised by the channel $N\to W_R^++e^-$. Such a fast decay
should have important effects on decoupling. It is this which is
studied in detail in the present paper. The effect on leptogenesis
has already received considerable attention \cite{ma}.

The effect of the decay of massive neutrinos on energy density,
nucleosynthesis, the Cosmic Microwave Background Radiation
(CMBR), galaxy formation, and stellar evolution has been
well-studied \cite{sato,dicus}. In these studies, the decay time
was taken to be large (typically, larger than 100-200 seconds),
and the effect of decay at the crucial epochs followed up.
The decay time was chosen greater than the decoupling time,
{\em i.e.}, it was assumed that the neutrinos first decoupled,
and, then, their cosmological and astrophysical effects were felt,
as a result of subsequent out-of-equilibrium decay \cite{dicus}.
If a fast decay like $N\to W_R^++e^-$ is considered, the
relationship of decoupling and decay may not be like this, and
the two have been considered, in the present paper, together, in
one Boltzmann equation.

The plan of the paper is as follows. Section 1 is the
Introduction. In section 2, the thermally averaged $N\bar N\to
W_R^+W_R^-$ annihilation cross-section times relative velocity
and the thermally averaged decay rate are calculated in the L-R
model, assuming pure Majorana neutrinos. The anomalous rate per
unit volume, used in ref.\cite{pa}, is slightly modified to
accommodate pure Majorana neutrinos. In section 3, the Boltzmann
equation is written down and treated approximately to obtain a
decoupling criterion. The possibility of decoupling is
investigated numerically. Section 4 discusses the conclusions.

\section{Annihilation, Decay, and Anomalous Reduction of
Right-handed neutrinos: Thermal averages in L-R model}

\subsection{Summary of relevant features of the L-R model}

In the L-R model, there are two doublets $ ( \nu,e_L)$ and $
(N,e_R)$ belonging to the representations $ ( \frac{1}{2} ,0,-1) $
and $ (0, \frac{1}{2} ,-1) $, respectively, of $ SU(2)_L\times
SU(2)_R\times U(1)_{B-L}$, where the quantum numbers refer to the
values of $ T_{L}, T_{R},$ and $B-L$ respectively. To simplify
the issues, only one generation is considered (the lightest), and
$N- \nu$, $W_L-W_R$, and $Z-Z'$ mixings are neglected.

The symmetry-breaking from $ SU(2)_L\times SU(2)_R \times
U(1)_{B-L} \to SU(2)_L\times U(1)_Y$ is achieved by means of a
scalar triplet ($T_{L}=0, T_{R}=1$ and $B - L = 2$) $\Delta \equiv
( \Delta^{++}, \Delta^+, \Delta^0)$, putting $ < \Delta^0> = v_R/
\sqrt{2}$ (chosen real). The right-handed gauge boson becomes
massive due to the piece of the Lagrangian
\begin{equation}
L_{\Delta W^+W^-}=D^\mu \Delta^\dagger
D_\mu \Delta
\label{eq:delww}
\end{equation}
with $ D_ \mu= \partial _ \mu + ig\vec{T}.\vec{W_\mu} + i(g^\prime/2)
B_ \mu$. From now on, the subscript $R$ is dropped except when
essential. The $T_i$ form a $3\times 3$ representation of
the $SU(2)$ generators in a spherical basis. $Z^\prime_\mu$ has the
definition
\begin{equation}
Z^\prime_\mu=( \sqrt {\cos2\theta}/{\cos\theta}
)W_3^\mu - \tan\theta B^\mu
\label{eq:zpr}
\end{equation}
(\ref{eq:delww}) gives a $W$ mass $
M_W^2=\frac{1}{2}g^2v_R^2$, and an interaction
\begin{equation}
L_{HW^+W^-}=\sqrt{2}gM_WW_\mu^+W^{\mu -}H
\label{eq:hww}
\end{equation}
where
$ H(x)/ \sqrt{2}=Re(\Delta^0(x))-v_R/ \sqrt{2}$, the displaced
neutral
Higgs field.
\\Neglecting $Z-Z'$ mixing, one gets the $Z'$ neutral current piece
\begin{equation}
L_{N\bar N}^{Z^\prime}=(g/2 \sqrt{\cos2\theta})
\cos\theta\bar N \gamma^\mu P_R N Z^\prime_\mu.
\label{eq:znn}
\end{equation}
$\theta$
is the weak mixing angle. The required charged current piece of the
Lagrangian is
\begin{equation}
L_{Ne}^W=(g/\sqrt{2})(\bar N \gamma^\mu P_R e W_\mu + h.c.).
\label{eq:wne}
\end{equation}
The $Z^\prime,W^+,W^-$ interaction piece is
\begin{eqnarray}
L_{W^+W^-}^{Z^\prime} & = & -ig( \sqrt{\cos2\theta}/\cos\theta)
[Z^{\prime\mu}
( \partial_\mu W_\nu^+W^{\nu-}-\partial_\nu W_\mu^+W^{\nu-}-
\partial_\mu W_\nu^-W^{\nu+} \nonumber \\
& & +\partial_\nu W_\mu^-W^{\nu+})
+\frac{1}{2}(\partial_\mu Z^\prime_\nu-\partial_\nu Z^\prime_\mu)
(W^{\mu +}W^{\nu -}-W^{\mu -}W^{\nu +})]
\label{eq:zww}
\end{eqnarray}

The Majorana mass of $N$ is thought to arise from the piece
\begin{equation}
L_{R\bar R\Delta}= f R^TC\epsilon( \vec{\tau}/ \sqrt{2}).
\vec{\Delta}R+h.c.
\label{eq:rrd}
\end{equation}
where $R^T$ is the $SU(2)_R$ doublet $(N,e_R)$, $C$ is the charge
conjugation matrix, and $\epsilon=i\tau_2$. $( \vec{\tau}/
\sqrt{2}).\vec{\Delta}$ is the matrix
\[
\left( \begin{array}{cc}
\Delta^+/\sqrt{2} & \Delta^{++} \\ \Delta^0 &
-\Delta^+/\sqrt{2}\end{array}\right) ,
\] and $\epsilon( \vec{\tau}/ \sqrt{2}).\vec{\Delta}$
is symmetric.
This term gives a Majorana mass for the $N$
neutrinos, $M= f v_R/\sqrt{2}$, so that $g/f=M_W/M=r_W$, and an
interaction
\begin{equation}
L_{N\bar NH}=(g/\sqrt{2}r_W) \bar N^CHN
\label{eq:nnh}
\end{equation}
$N^C$ is the field conjugate to $N$. It has been pointed
out \cite{flan,sar} that the effective one-loop mass matrix for
multi-generation unstable Majorana neutrinos is not Hermitian
and, strictly speaking, $N\not= N^C$. But, here, at the level of
tree-order calculations for a single generation, $N=N^C$ will be
assumed.

\subsection{$<\sigma |v| >$ for $N\bar N\to W_R^+W_R^-$}

Three amplitudes $iM_{Z^\prime},iM_e,iM_H$ have been considered
for $N\bar N\to W^+W^-$, these arising, respectively, from $N\bar
N\stackrel{Z^\prime}{\to}W^+W^-$ in the s-channel, $N\bar
N\stackrel{e}{\to}W^+W^-$ in the t-channel, and $N\bar
N\stackrel{H}{\to}W^+W^-$ in the s-channel. Earlier calculations,
with $r_W \ll 1$ \cite{6}, considered a heavy charged lepton
exchange in the t-channel. In this paper, as required by the L-R
model, an ordinary electron is considered to be exchanged.

$iM_{Z^\prime}$ is calculated from (\ref{eq:znn}) and
(\ref{eq:zww}), $iM_e$ from (\ref{eq:wne}), and $iM_H$ from
(\ref{eq:hww}) and (\ref{eq:nnh}). One finds that, in the limit
$M\to 0$ and $s\gg M_W^2$, $iM_e$ and $iM_{Z^\prime}$ cancel in
tree-order, as expected. With a massive $M\gg M_W$, an extra
$M$-dependent term remains in $iM_{Z^\prime}+iM_e$. On calculating
$\sum|{\cal M} |^2$, it is found that the interference terms
between $iM_H$, and $iM_{Z^\prime}+iM_e$ cancel, and for $r_W \ll
1$, in the CM frame,
\begin{equation}
\sum|{\cal M}|^2=\frac{2 g^4}{r_W^4}\left[\frac{1}{t^2}E^2{\bf
k}^2\sin^2\theta_{CM}+\frac{16}{M^2(s-M_H^2)^2}E^6\right].
\label{eq:matz+esq}
\end{equation}
$(E,\vec{\bf k})$ is the 4-momentum of the $N$ neutrino in the CM
frame, and $\theta_{CM}$, the angle of scattering.
$\bf k$ has been written for $|\vec{\bf k}|$.

The thermally
averaged cross-section times relative velocity is
\begin{eqnarray}
<\sigma|v|> &=& (1/n_{eq}^2)\int d\pi_Nd\pi_{\bar
N}d\pi_{W^+}d\pi_{W^-}(2\pi^4)\delta^4(p_N+p_{\bar
N}-p_{W^+}-p_{W^-})\nonumber \\
& &\times \sum|\bar {\cal
M}|^2e^{-E_N/T}e^{-E_{\bar N}/T}.
\label{eq:sigv}
\end{eqnarray}
$n_{eq}$ is the equilibrium value of the number density $n$ of
the $N$ neutrinos.
\begin{equation}
n_{eq}=g_N\left[\frac{MT}{(2\pi)}\right]^{\frac{3}{2}}e^{-M/T}.
\label{eq:neq}
\end{equation}
$g_N=2$ for Majorana neutrinos. The measure
$d\pi_i=g_id^3p_i/[(2\pi)^32E_i].$ $\sum|\bar {\cal M}|^2$ is the
spin-averaged matrix element squared, with symmetry factor 1/2!,
arising from the identification $N=N^C$.

The invariant integral $\int
d\pi_{W^+}d\pi_{W^-}(2\pi)^4\delta^4(p_N+p_{\bar
N}-p_{W^+}-p_{W^-})\sum|\bar {\cal M}|^2$ is calculated in the CM
frame, then transformed to the comoving ``lab" frame, in which
$N,\bar N$ have energies $E_N,E_{\bar N}$, respectively, and
thermally averaged according to (\ref{eq:sigv}). In all
calculations, $N$ neutrinos are non-relativistic, {\em viz.}
$E_N=M+{\bf k}_N^2/(2M),\; {\bf k}_N^2\ll M^2.$ This is because
the interesting region for decoupling studies of a massive
particle has $T\ll M.$

The result is
\begin{equation}
<\sigma|v|> =
\frac{1}{2!}\frac{g^4}{64\pi r_W^4}\left[
\frac{T}{M^3} +
\frac{16}{ M^2(4-r_H^2)^2}\left\{1- \frac{3 T (4+r_H^2)}{2 M
(4-r_H^2)} \right\} \right].
\label{eq:sigv2}
\end{equation}
$r_H=M_H/M$.\\
The first term arises from $Z'$- and $e$-exchange, and apart from
the 1/2! factor, agrees with the result of \cite{8}, where the
results of \cite{6} have been considered in the limit $s\to
4M^2$. The second term originates from $H$-exchange. In the
calculation of this term, a further approximation has been made,
{\em viz.} $4{\bf k}^2\ll |4M^2-M_H^2|$, {\em i.e.}, this
calculation is reliable provided $M_H$ is not very near in value
to 2$M$. In \cite{6}, the $H$-exchange contribution was found to be
negligible as $s\to 4M^2$, because of a factor $(s-4M^2)$ which
arises for Dirac neutrinos.  For Majorana neutrinos, this factor
is absent, and this term cannot be neglected. As noted
earlier, there is no interference between the $H$-exchange
amplitude and those arising from $Z'$- and
$e$-exchange. In \cite{pa}, only the first term in
(\ref{eq:sigv2}) was considered, which is the approximation
$M_H\gg M$.

The processes $N\bar N\to F\bar F$, where $F$ is a relevant
fermion, have not been considered here. For $M\gg M_W$, the
contribution of these processes is small compared to that of
$N\bar N\to W^+W^-$ \cite{6}, and their effect on decoupling is
overshadowed by the effects of $N\bar N\to W^+W^-$ and the
anomalous reduction of $N$ neutrinos \cite{pa}.

\subsection{Thermally averaged decay width of $N$ neutrinos}

The decay width for $N\to W^++e^-$ can be calculated from
(\ref{eq:wne}). Calculation gives the spin-averaged matrix
element squared in the neutrino rest frame
\[ \sum |
\bar {\cal M}|^2 = g^2\left[ \frac{1}{2}M{\bf p}+\frac{M(M-{\bf
p})M{\bf p}}{M_W^2} \right], 
\]
with ${\bf p}=\frac{1}{2}M(1-r_W^2)$ being the momentum of the
decay products, resulting in the width
\begin{equation}
\Gamma_e=\frac{g^2 M}{32\pi}(1-r_W^2)^2\left(1+\frac{1}{2r_W^2}\right)
.
\label{eq:gamme}
\end{equation}
In the frame in which the neutrino has energy $E_N$, the width
becomes
\begin{equation}
\Gamma_e^E=(M/E_N)\Gamma_e.
\label{eq:gammeE}
\end{equation}
The
thermally averaged decay width \cite{19,26} is defined as
\begin{eqnarray} \bar
{\Gamma}_e & = & (1/n_{eq})\int
d\pi_Nd\pi_{W^+}d\pi_{W^-}(2\pi)^4\delta^4(p_N-p_{W^+}-p_{W^-})
\sum | \bar {\cal M}|^2e^{-E_N/T} \nonumber \\ & = & (1/n_{eq})\int
(g_Nd^3p_N/(2\pi)^3)\Gamma _e^Ee^{-E_N/T}. \nonumber \end{eqnarray}
Using (\ref{eq:gamme}) and (\ref{eq:gammeE}), one gets
\begin{equation}
\bar{\Gamma}_e=\frac{g^2 M}{32\pi
}(1-r_W^2)^2\left(1+\frac{1}{2r_W^2}\right)
\left(1-\frac{3T}{2M}\right).
\label{eq:gammebar}
\end{equation}
The calculation has been done in the approximation $T\ll
M,{\bf k}_N^2\ll M^2.$ For $M\gg M_W^2$,
\begin{equation}
\bar{\Gamma}_e=\frac{g^2 M}{64\pi r_W^2}\left(1- \frac{3T}{2M}\right).
\label{eq:gammebar2}
\end{equation}

If the mass of the $N$ neutrino is greater than the $\Delta^+$
mass, (\ref{eq:rrd}) allows the decay $N\to \Delta^++e^-$. The
corresponding
\[ \sum |\bar {\cal M }|^2
=(1/2)f^2(M^2-M_+^2),\]
where $M_+$ is the $\Delta^+$ mass. In the
rest frame of the neutrino, this decay width comes out as
\begin{equation}
\Gamma_+=(g^2/64\pi)(M^2/M_W^2)M(1- r_+^2)^2,
\label{eq:gamm+}
\end{equation}
where $r_+ = M_+/M$. There is another Yukawa piece of the Lagrangian
\begin{equation}
L_{LR\Phi}
=\sum_{i,j}(h_{ij}\bar {L}_i\Phi R_j+
h^\prime_{ij}\bar
{L}_i\tau_2\Phi^\star\tau_2R_j)+h.c.
\label{eq:lyuk2}
\end{equation}
The scalar field $\Phi$ transforms under the gauge group as
$(\frac{1}{2},\frac{1}{2},0)$ and is represented by the matrix
\[\Phi=\left( \begin{array}{cc}\phi^0 & \phi^{\prime +}\\
\phi^- & \phi^{\prime 0} \end{array}\right). \]
Such pieces have been used in different models to break CP and
induce baryogenesis through
leptogenesis \cite{fuku,pil,luty,vay,flan,ma}. If one generation is
considered, the coupling constant $h$ contributes to the electron
mass and must be very small. To fit the observed baryon
asymmetry, with $N$ mass in the TeV range, $|h_{ij}|^2$ values of
the order of $10^{-10}$ to $10^{-13}$ have been
considered \cite{fuku,pil}. In this situation, as CP breaking has
not been considered here, and only one generation taken into
account, it has not been thought useful to consider $N$-decays
arising from (\ref{eq:lyuk2}). In any case, it will be found that
any enhancement of the decay width given in (\ref{eq:gammebar})
and (\ref{eq:gammebar2}) will only strengthen the main conclusion
of the paper.

\subsection{Anomalous generation of Majorana neutrinos}

The one generation, sphaleron-mediated, fermion number violating
transition rate per unit volume, with $|\Delta L|=1$, $|\Delta
B|=1,$ for the quantum anomaly, involving the $SU(2)_R$ gauge
group, was written, in \cite{pa}, by extrapolation from the
$SU(2)_L$ case \cite{25}, as
\begin{eqnarray}
A_R & = & (1.4\times 10^6) \left(\frac{ b
M_W^7}{g^6T^3}\right)\left[1-\left(\frac{T}{zM_W}\right)^2\right]^{7/2}
\nonumber \\ & & \times \exp\left\{-\frac{16\pi
M_W}{g^2T}\left[1-\left(\frac{T}{zM_W}\right)^2\right]^{1/2}\right\}.
\label{eq:ar}
\end{eqnarray}
$M_W$ is the zero temperature $W_R$ mass. $z=T_R/M_W$, where $T_R$
is the critical temperature associated with the breaking of the
$SU(2)_R$ gauge symmetry. So, $z$ is essentially a
quantity which reflects the uncertainty in the values of the L-R
model parameters, while $b$ captures the uncertainties
involved in the extrapolation from the $SU(2)_L$ to the $SU(2)_R$
case in addition to those in the estimation of the prefactor of
the anomaly driven transition \cite{moore}.

In \cite{pa}, the anomalous rate of reduction of $N$
neutrinos was considered, maintaining a distinction between $N$ and
$N^C$. Here, $N=N^C$ neutrinos are considered, to maintain uniformity
with the $N\bar N\to W^+W^-$ calculations. This entails an extra
factor of 2, as seen below.

For an anomalous process $l$, with
$\Delta_L=+1$, such that
\[l:i+j+\cdots\to N+a+b+\cdots,\]
one writes \cite{pa},
\begin{eqnarray}
A_l & = & \int d\pi_Nd\pi_ad\pi_b\cdots
d\pi_id\pi_j\cdots|{\cal M}_l|^2 \nonumber \\
& & \times
(2\pi)^4\delta^4(p_N+p_a+p_b+\cdots-p_i-p_j-\cdots)f^{eq}_Nf_af_b
\cdots \nonumber \\
& = & I_ln_{eq},
\label{eq:al}
\end{eqnarray}
where $I_l$ contains the result of the
phase space integrations, apart from $n_{eq}$ \cite{19,26}, and
$i,j, \cdots a,b, \cdots$ are all supposed to be in equilibrium.

Taking the view that leptogenesis and baryogenesis are effects of a
smaller order, CP-symmetry is assumed. Then \cite{pa}, for each
process $l$, there is a $\Delta L=-1$ process
\[\bar
l^{\prime}:N+a+b+\cdots\to i+j+\cdots , \]
with the same $|{\cal M}_l|^2$. For this process,
\begin{eqnarray}
A_{\bar l^{\prime}} & = & \int
d\pi_Nd\pi_ad\pi_b\cdots d\pi_id\pi_j\cdots|{\cal M}_l|^2
\nonumber \\
& & \times(2\pi)^4\delta^4(p_N+p_a+p_b+\cdots-p_i-p_j-\cdots)f_Nf_
af_b\cdots \nonumber \\
& = & I_ln
\label{eq:al2}
\end{eqnarray}
The $\Delta L=+1$ processes, generating $N$
neutrinos, add up to
\[\sum_lA_l=n_{eq}\sum_lI_l=\frac{1}{2}A_R\exp(-\beta\mu_L/2),\]
from \cite{25}. $\mu_L=\mu_N$ is the chemical potential
($\mu_e=0$, as electrons are in equilibrium). The 1/2 factor
arises from the assumption \cite{pa}, that, to a first
approximation, the rate of generation of one member of a lepton
doublet may be taken to be the same as that of the other, near
equilibrium (not a bad condition at decoupling when the neutrinos
are just falling out of equilibrium).\\Similarly,
\[
\sum_{\bar l^\prime}A_{\bar
l^\prime}=n\sum_lI_l=\frac{1}{2}A_R\exp(+\beta\mu_L/2).\]
So, for small $\mu_N$ \cite{25},
\[
\begin{array}{c}n_{eq}\sum_lI_l\approx\frac{1}{2}A_R(1-\beta\mu_N/2)
,\\n\sum_lI_l\approx\frac{1}{2}A_R(1+\beta\mu_N/2).\end{array}
\]
One gets
\[\sum_lI_l=A_R/(n+n_{eq}),\]
and the anomalous rate of
reduction of $N$ neutrinos per unit volume
\begin{equation}
A_N=\sum_{\bar
l^\prime}A_{\bar l^\prime}-\sum_lA_l=\;\frac{
n-n_{eq}}{ n+n_{eq}}A_R.
\label{eq:an}
\end{equation}
This has an extra
factor of 2, compared to \cite{pa}, because anti-particle processes,
which had to be considered separately there, do not appear here,
because of the assumption $N=N^C$.

\section{Effect of decay on decoupling}

\subsection{The Boltzmann equation for $N$ neutrinos}

Using the results of the last section, one can
write the Boltzmann equation
\begin{equation}\frac{
dn}{ dt}+3Hn=-2<\sigma|v|>(n^2-n_{eq}^2) - 2\bar
\Gamma_e(n-n_{eq})-\frac{ n-n_{eq}}{
n+n_{eq}}A_R,
\label{eq:bolt}
\end{equation}
where the second term on the
left gives the effect of expansion, and the three terms on the right
are to be taken from (\ref{eq:sigv2}), (\ref{eq:gammebar2}), and
(\ref{eq:an}), respectively.\\The expression for the Hubble
parameter $H$ is
\begin{equation}
H=1.66g^{\star \frac{1}{2}}T^2/M_{Pl}.
\label{eq:hubb}
\end{equation}
$g^\star$ is taken $\approx 100$, and $M_{Pl}=1.22\times 10^{19}$
GeV. The 2 factor with $<\sigma|v|>$ appears because two neutrinos
are disappearing in $N\bar N\to W^+W^-$, considering $N=N^C$. The 2
factor with $\bar \Gamma_e$ appears because of the two decay channels
$N\to W^++e^-$ and $N\to W^-+e^+$.Writing $x=M/T$ and $Y=n/s$,
where $s=g^{\star S}(2\pi^2/45)T^3,\quad\mbox{with} \quad g^{\star
S}\approx 100,$ (\ref{eq:bolt}) becomes
\begin{equation}
\frac{ dY}{
dx}=-f(x)(Y^2-Y_{eq}^2)-d(x)(Y-Y_{eq})-g(x)\frac{
Y-Y_{eq}}{ Y+Y_{eq}}.
\label{eq:bolt2}
\end{equation}
In (\ref{eq:bolt2}),
\begin{equation}
f(x)=\left(\frac{1.41\times 10^{16}{\rm
~GeV~}}{M_W}\right)\left(\frac{a}{x}\right)^3\left[\frac{2x}{(1-r_H^2/4)^2}
- h(r_H) \right],
\label{eq:f}
\end{equation}
where $a = 1/r_W = M/M_W$ and 
\begin{equation}
h(y)=\frac{1}{(1 - y^2/4)^3}\left[ 1 +
y^2\left\{(1-y^2/8)(1-y^2/4)+5/4\right\}\right] ,
\label{eq:h}
\end{equation}
\begin{equation}
d(x) = \left(\frac{3.08 \times 10^{15}{\rm ~GeV~}}{M_W}\right) 
a x \left[ 1 - \frac{3}{2x} \right], 
\label{eq:d}
\end{equation}
\begin{eqnarray}
g(x) & = & \left(\frac{3.06 \times
10^{23}{\rm ~GeV~}}{M_W}\right) \frac{b}{a} 
\left[ \frac{1}{u^2} - \frac{1}{z^2}\right]^\frac{7}{2} \nonumber
\\ & & \times \exp\left\{-118.98 \left[ \frac{1}{u^2} -
\frac{1}{z^2}\right]^\frac{1}{2}\right\},
\label{eq:g}
\end{eqnarray}
with $u = a/x = T/M_W$.

Writing $Y=Y_{eq}+\Delta$, it is noted that, before decoupling, $Y$
is close to $Y_{eq}$, and $\Delta^\prime$ may be put equal to
zero. Then, (\ref{eq:bolt2}) may be put in the form
\[
\Delta=\frac{ -Y_{eq}^\prime}{
f(x)(2Y_{eq}+\Delta)+d(x)+\frac{ g(x)}{
2Y_{eq}+\Delta}}.
\]

From $n_{eq}$ in (\ref{eq:neq}), one gets
\[
Y_{eq}=2.89\times 10^{-3}x^\frac{3}{2}e^{-x},\quad \mbox{and}\quad
Y_{eq}^\prime\approx -Y_{eq},
\]
at decoupling, when it is expected that $x=x_d \gg 1$.
The criterion for decoupling may be taken as $\Delta=c^\prime
Y_{eq}$, where $c'$ is of order 1. As in \cite{pa}, $c'$ is chosen to
be 1. \\Then, the condition for decoupling is
\begin{equation}
3f(x_d)Y_{eq}(x_d)+d(x_d)+\frac{
g(x_d)}{ 3Y_{eq}(x_d)}=1
\label{eq:decoup}
\end{equation}

This is the key condition in our analysis of decoupling. Each
factor on the lhs of (\ref{eq:decoup}) is positive \cite{fpos}.
So, there is no scope for cancellation between different terms.
We will show in the following subsections that, in fact, the
condition can never be satisfied. The argument proceeds as
follows. First, we consider (\ref{eq:decoup}), excluding the
second (decay) term in the lhs. We show that there is a value
$x_a$, which we obtain numerically below, such that, for $x <
x_a$ or $x > x_a$, the lhs is, respectively, greater or less than
unity in the absence of decay. We then check that for $x > x_a$
the decay term is much larger than unity so that there is no
value of $x$ (= $M/T$) for which eq.  (\ref{eq:decoup}) is
satisfied.

\subsection{Decoupling in the absence of decay}

First, $d(x_d)$ is omitted, and the decoupling condition 
\begin{equation}
l(x)= 3f(x)Y_{eq}(x)+\frac{g(x)}{ 3Y_{eq}(x)}=1
\label{eq:nodec}
\end{equation}
is solved to give $x=x_a$. $x_a$ represents the value of $M/T$ for
which decoupling would occur in the absence of decay. 

$Y_{eq}$ gives a factor $e^{-x}$, and, so, the term with $f(x)$
increases as $x$ decreases. $g(x)$ has an exponential factor of
the form $e^{-E_{sp}/T}=e^{-Kx/a}$ \cite{pa,16,25,32,33}, where
$E_{sp}$ is the energy of the sphaleron mode which decays to
cause anomalous $L$ generation. The kinematic constraint on $N$
production, $E_{sp}>M,$ gives $K/a>1\; \mbox{\cite{pa}, and
so}\quad g(x)/Y_{eq}\sim e^{-(\frac{K}{a}-1)x}$, and, again,
increases as $x$ decreases. This means that $l(x)>1$ if $x<x_a$.
So, $x<x_a$ will not satisfy the decoupling condition
(\ref{eq:decoup}). Therefore, on the whole, it may be said
approximately \cite{19} that, for $x>x_a$, the annihilation rate
plus the anomalous reduction rate is less than $H$.

This programme is followed numerically. First, for definiteness,
we fix $b$ and $z$, the two parameters in the expression for
$g(x)$ -- see eq. (\ref{eq:g}) -- at the values 1 and 4,
respectively \cite{pa}. $a$ is varied from 2-100, for
$r_H$\footnote{$h(r_H)$ assumes its asymptotic value of --2 for
$r_H >$ 5.} = 0, 1, 3, 10 and $x_a$ found by solving
({\ref{eq:nodec}) numerically.

\begin{table}[htb]
\begin{center}
\begin{tabular}{|l|r| l|r| l|r|}
\hline
\multicolumn{2}{|c|}{$r_H$=1} & \multicolumn{2}{|c|}{$r_H$=3} &
\multicolumn{2}{|c|}{$r_H$=10} \\ \hline $a$ &
\multicolumn{1}{|c|}{$x$} & $a$ & \multicolumn{1}{|c|}{$x$} & $a$ &
\multicolumn{1}{|c|}{$x$}\\ \hline
2 & 25.79 & 2 & 25.06 & 2 & 22.30 \\ \hline
5 & 28.49 & 5 & 27.74 & 5 & 24.89\\ \hline
10 & 30.54 & 10 & 29.77 & 10 & 26.86\\ \hline
20 & 32.60 & 20 & 31.81 & 20 & 28.83\\
\hline 50 & 35.32 & 50 & 34.52 & 50 & 32.18\\
\hline 75 & 73.77 & 75 & 73.77 & 75 & 73.77\\ \hline 100 & 238.83
& 100 & 238.83 & 100 & 238.83 \\ \hline
\end{tabular}
\vskip 20pt
{\small \sf Table I: $x = M/T$ at decoupling for different
choices of $a = M/M_W$ and $r_H = M_H/M$ without the inclusion of
the decay contribution in the Boltzmann equation. $M_W$ has been
chosen to be 4000 GeV}
\end{center}

\end{table}

The results for $M_W=4000$ GeV are shown in Table I for
$r_H$=1,3,10. There is a clear trend, showing an increase in
$x_a$ for an increase in $a$ (constant $r_H$), and a decrease in
$x_a$ for an increase in $r_H$ (constant $a$). The values for
$r_H$=100 and $r_H$=1000, we have checked, do not differ at all.It is safe to say that $x_a > 20$ for the parameter ranges considered.

\begin{table}[htb]
\begin{center}
\begin{tabular}{|c|r| c|r|}
\hline
\multicolumn{2}{|c|}{$a$=2} & \multicolumn{2}{|c|}{$a$=50} \\
\hline $b$ &
\multicolumn{1}{|c|}{$x$} & {$b$} & \multicolumn{1}{|c|}{$x$}\\ \hline
$10^3$ & 22.30 & $10^3$ & 36.79\\
\hline $10^2$ & 22.30 & $10^2$ & 35.11\\
\hline 10 & 22.30 & 10 & 33.50\\ \hline 1 & 22.30 & 1 & 32.18\\ \hline
$10^{-1}$ & 22.30 & $10^{-1}$ & 31.60\\ \hline $10^{-2}$ & 22.30 &
$10^{-2}$ & 31.47\\ \hline $10^{-3}$ & 22.30 & $10^{-3}$ & 31.46\\
\hline
\end{tabular}
\vskip 20pt
{\small \sf Table II: Effect of uncertainty in anomalous rate on
the results of Table I}
\end{center}\end{table}

We now address the uncertainty regarding the anomalous rate
\cite{moore}. The uncertainty is embodied in the parameters $b$
and $z$ appearing in the expression for this rate.  The parameter
$z=T_R/M_W$ is of order unity ($z=3.8$ in the $SU(2)_L$ case
\cite{25}).  To gauge the sensitivity of the results on $z$, a
calculation was made with $a$=50 (to give good weightage to the
anomalous transition factor $g(x)$) and $r_H$=1.  $x_a$, obtained
by solving (\ref{eq:nodec}), was 42 for $z=2$, 35.6 for $z=3$,
and reached an asymptotic value of 35.3 for $z\ge 4$. For
$r_H$=10, $x_a$ was 42 for $z=2$, 35 for $z=3$, and had an
asymptotic value of 31.5. The conclusion $x_a > 20$, of the last
paragraph is not disturbed.  In \cite{pa}, the uncertainty in $b$
was taken into account by varying it through six orders of
magnitude about $b=1$. We do a similar calculation for $z=4$,
$r_H=10$, $M_W=4000$ GeV and present the results in Table II. It
is clear that even this large variation of $b$ through six orders
of magnitude does not affect the conclusion of the previous
paragraph.

For comparison, the results for $M_W=2000$ GeV, for the
same values of the other parameters, are shown in Table III.

\begin{table}[htb]
\begin{center}
\begin{tabular}{|l|r| l|r| l|r|}
\hline
\multicolumn{2}{|c|}{$r_H$=1} & \multicolumn{2}{|c|}{$r_H$=3} &
\multicolumn{2}{|c|}{$r_H$=10} \\ \hline $a$ &
\multicolumn{1}{|c|}{$x$} & $a$ & \multicolumn{1}{|c|}{$x$} & $a$ &
\multicolumn{1}{|c|}{$x$}\\ \hline
2 & 26.47 & 2 & 25.73 & 2 &
22.95\\ \hline 5 & 29.18 & 5 & 28.41 & 5 & 25.54\\ \hline
10 & 31.23 & 10 & 30.45 &
10 & 27.51\\ \hline
20 & 33.28 & 20 & 32.49 & 20 & 29.49\\ \hline
50 & 36.00 & 50 & 35.20 & 50 & 32.75\\ \hline
75 & 75.01 & 75 & 75.01 & 75 & 75.01\\ \hline
100 & 242.83 & 100 & 242.83 & 100 & 242.83 \\ \hline
\end{tabular}
\vskip 20pt
{\small \sf Table III: Same as in Table I but for $M_W$ = 2000 GeV}
\end{center}

\end{table}

The expectation that $l(x)$ -- see eq. (\ref{eq:nodec}) --
decreases below 1, as $x$ increases through $x_a$, was verified
numerically for $M_W=4000$ GeV, and $a$ = 10,20,50,75, for each
of $r_H$=1,10. Table IV shows the results for $r_H$ = 1,10, and
$a$ = 10,50.

\begin{table}[htb]
\begin{center}
\begin{tabular}{|c|r|c|r|c|r|c|r|} \hline
\multicolumn{4}{|c}{$r_H$=1} & \multicolumn{4}{|c|}{$r_H$=10}
\\ \hline
\multicolumn{2}{|c|}{$a$=10} & \multicolumn{2}{c}{$a$=50} &
\multicolumn{2}{|c}{$a$=10} & \multicolumn{2}{|c|}{$a$=50}
\\ \hline
$x$ & \multicolumn{1}{|c|}{$l(x)$} &
$x$ & \multicolumn{1}{|c|}{$l(x)$} &
$x$ & \multicolumn{1}{|c|}{$l(x)$} &
$x$ & \multicolumn{1}{|c|}{$l(x)$} \\ \hline
20 &
45.23 $\times 10^3$ & 30 & 227.87 & 20 & 14.61 $\times 10^2$ & 25 &
13.13 $\times 10^3$\\ \hline 35 & $1.09 \times 10^{-2}$ & 40 &
8.70 $\times 10^{-3}$& 30 &
3.67 $\times 10^{-2}$ & 40 & 1.51 $\times 10^{-4}$\\ \hline
\end{tabular}
\vskip 20pt
{\small \sf Table IV: Behavior of $l(x)$ (\ref{eq:nodec}) around
$x_a$ in the absence of decay. $M_W$ = 4000 GeV for this Table.}
\end{center}
\end{table}

\subsection{Effect of decay}

From the numerical results of the previous subsection it would be
safe to say that for $x<20, l(x)>1$. Now, if one looks at $d(x)$
(\ref{eq:d}), it is clear that, for $x>20, d(x)\gg 1$, and,
moreover, as $x$ increases further, $d(x)$ increases. So, there
is no possibility of the decoupling condition (\ref{eq:decoup})
being satisfied.

$d(x)$, of course, has a simple physical meaning. In the lhs of
(\ref{eq:bolt}), 
\[ \frac{ dn}{
dt}+3Hn=Hsx\frac{ dY}{ dx}.
\] 
Comparing
(\ref{eq:bolt}) and (\ref{eq:bolt2}), $d(x)=(2\bar \Gamma_e/Hx).$
If $d(x)\gg 1$ for $x=x_a$, with $x_a\sim 20$, this means that
$\bar \Gamma_e\gg H$ at this point. Considering the physical
meaning of $x_a$, it may be concluded that, although, at
temperatures lower than $T_a=M/x_a$, the annihilation rate plus
anomalous reduction rate falls below the expansion rate, the
decay rate remains much faster than the expansion, and this
prevents decoupling.

In the absence of decoupling, the fast decay constrains the $N$
neutrino number density to follow the equilibrium density $\sim
e^{-M/T}$.  There is no out-of-equilibrium decay.

One may check that this conclusion is not an artefact of the
approximation $M^2\gg M_W^2$ in the calculations. Without this
approximation, (\ref{eq:gammebar}) gives the decay rate
\begin{equation}
\bar\Gamma_e=\frac{
g^2M(a^2-1)^2(a^2+2)(1-(3/2x))}{ 64\pi a^4}
\end{equation}
so that
\begin{equation}
d(x)=\left(\frac{ 3.08\times
10^{15}{\rm
~GeV~}}{M_W}\right)\frac{(a^2-1)^2(a^2+2)x}{a^5}\left[1 -
\frac{3}{2x}\right].
\end{equation}
Simple calculation shows that if $d(x)$ is to be $<1$,
for $x\sim 20,$ then one must have $a\sim 1$, to 1 part in
$10^7$ (for $M_W=4000$ GeV). There is no reason for such fine
tuning between $N$ and $W$ masses.

If the $N$ decay width is augmented by the
$\Delta^+,e^-$ channel, {\em i.e.}, if $M > M_+$, 
\begin{equation}
d(x)=\left(\frac{3.08\times 10^{15}{\rm ~GeV~}}{M_W}
\right)\left[\frac{(a^2-1)^2(a^2+2)}{ a^5}+a(1-r_+^2)^2 
\right]x \left[1 - \frac{3}{2x}\right].
\end{equation}
In this case, if $d(x)$ is to be $<1$ for $x\sim 20$, with
$M_W=4000$ GeV, not only must $r_W\sim 1$, but, also, $r_+\sim 1$,
each to 1 part in $10^7$, an unacceptable situation.

Clearly, if the decay width is further augmented by introducing
other channels, this will change $ \Gamma_e$, but not the
$x$-dependence of $\bar \Gamma_e$ or $d(x)$, so that $d(x)<1$ for
$x\sim20$ will be an even remoter possibllity.

$d(x)$ can be decreased by increasing $M_W$. Taking $r_H=1$, a
little calculation shows that $d(x)=0.999$ and $l(x)=0.001$ for
$x=4.25$ and $M_W=1.7\times 10^{16}$ GeV if $a=2$, and for
$x=6.6$ and $M_W=1.6\times 10^{17}$ GeV if $a=10$.

This scale of $M_W$ agrees with \cite{ma} where the decoupling
condition was, however, chosen simply as $\bar\Gamma_e<H$ at $T=M$,
$\bar\Gamma_e$ having been assigned the value $g^2T/8\pi$. In
this paper, we have calculated $\Gamma_e=\frac{
g^2(M^2-M_W^2)^2(1+M^2/2M_W^2)} { 32\pi M^3}$, and taken the
thermal average of $(M/E_N)\Gamma_e$ for $T<M$. We have used a
decoupling criterion which includes the effect of annihilations
and anomalous reduction, for a wide range of the parameters
$r_H=M_H/M$ and $r_W=M_W/M$. Also, our method of approximate
solution of the Boltzmann equation keeps the temperature of
decoupling open and calculable, {\em e.g.}, in the previous
paragraph it was found that decoupling occurs for $M>10^{16},
(10^{17})$ GeV at $T=M/4$ ($M/6$), and not at $T=M$.

So, the main point which emerges is that for there to be
decoupling of massive neutrinos in the L-R model, it is necessary
to consider right-handed gauge boson mass values far above the
physically expected L-R mass scale. Further as $M=aM_W$, the
see-saw mechanism will, then, give values of the $\nu$ mass,
which will be unacceptably small, when compared to neutrino
oscillation values of $\Delta m^2$.

\section{Conclusions}

It is to be concluded that, in the L-R model, with right-handed
neutrino mass greater than the $W_R$ mass, even when the
annihilation plus anomalous reduction rate for these neutrinos
has fallen below the expansion rate, the decay remains faster
than expansion, and becomes increasingly faster. So, massive
right-handed neutrinos will decay while remaining in equilibrium.

We find that this is true for a wide range of values of $M_W/M$ and
$M_H/M$, and also for a wide allowance of uncertainty in the
anomalous rate.

As the equilibrium number density varies as 
$e^{-M/T}$, right-handed neutrinos rapidly dwindle in number
once the temperature falls below their mass. The decay products, with
much lower masses, equilibriate immediately. Hence, there is no
question of influencing the present density of the universe, CMBR,
and nucleosynthesis or later events, and no mass bound can be set
for right-handed neutrinos in the L-R model, if their mass is greater
than the $W_R$ boson mass (apart from an upper bound from
unitarity \cite{7,8}).

Because there is no
out-of-equilibrium decay, the lepton (and baryon) number generation
scenarios, utilising the decay of massive Majorana neutrinos will not
work in the L-R model, if the neutrino mass is greater than the $W_R$
boson mass, a result also previously noted in \cite{ma}.

\vskip 20pt

\parindent 0pt

{\large {\bf Acknowledgement:}} The research of AR has been
supported in part by a grant from C.S.I.R., India. PA wishes to
thank Dr. S. Datta of Presidency College, Calcutta, for permission
to use facilities.


\begin{thebibliography}{99}
\bibitem{fpos} At first sight it might seem from eq. (\ref{eq:f})
that the first term on the rhs of eq. (\ref{eq:decoup}) is not
always positive. That this is not the case can be seen by
noting that (a) this analysis is valid only when $r_H$ is away
from the value 2, and, from eq. (\ref{eq:h}), (b) $h(r_H)$ is
negative for $r_H > 2$ and quickly assumes its asymptotic value of
--2 for $r_H > 5$, and (c) for $r_H \le 1$, though $h(r_H)$ is
positive, the first term in the square brackets in eq.
(\ref{eq:f}) is dominant for $x\gg 1$ (expected at decoupling).
\end{thebibliography}
\end{document}